\documentclass[conference]{IEEEtran}

\IEEEoverridecommandlockouts
\usepackage{cite}
\usepackage{amsmath,amssymb,amsfonts}
\usepackage{algorithmic}
\usepackage{graphicx}
\usepackage{multirow}
\usepackage{textcomp}
\usepackage{xcolor}
\usepackage{xspace}
\usepackage{listings}
\usepackage{hyperref}
\usepackage{geometry}

\lstdefinelanguage{Fstar}{
  morekeywords={module, let, type, val, assume, begin, end, function, forall, fun, inline_for_extraction},
  sensitive=true,
  morecomment=[l]{//},
  morecomment=[s]{(*}{*)},
  morestring=[b]",
}

\def\BibTeX{{\rm B\kern-.05em{\sc i\kern-.025em b}\kern-.08em
    T\kern-.1667em\lower.7ex\hbox{E}\kern-.125emX}}
\begin{document}
\newgeometry{top=72pt, bottom=54pt, left=54pt, right=54pt}

\title{UAV Resilience Against Stealthy Attacks}
 
\author{\IEEEauthorblockN{Arthur Amorim}
\IEEEauthorblockA{\textit{Department of Computer Science} \\
\textit{University of Central Florida\\}
Orlando, USA \\
arthur.amorim@ucf.edu}
\and
\IEEEauthorblockN{Max Taylor}
\IEEEauthorblockA{\textit{Research Acceleration Department} \\
\textit{Idaho National Laboratory}\\
Idaho Falls, USA \\
maxwell.taylor@inl.gov}
\and
\IEEEauthorblockN{Trevor Kann}
\IEEEauthorblockA{
\textit{Electrical and Computer Engineering}
\textit{Carnegie Mellon University}\\
Pittsburgh, USA \\
tkann@cmu.edu}
\and
\IEEEauthorblockN{Gary T. Leavens}
\IEEEauthorblockA{\textit{Department of Computer Science} \\
\textit{University of Central Florida}\\
Orlando, USA \\
leavens@ucf.edu}
\and
\IEEEauthorblockN{William L. Harrison}
\IEEEauthorblockA{\textit{Research Acceleration Department} \\
\textit{Idaho National Laboratory}\\
Idaho Falls, USA \\
william.harrisonr@inl.gov}
\and
\IEEEauthorblockN{Lance Joneckis}
\IEEEauthorblockA{\textit{Research Acceleration Department} \\
\textit{Idaho National Laboratory}\\
Idaho Falls, USA \\
lance.joneckis@inl.gov}
}

\author{}
\author{Arthur~Amorim\IEEEauthorrefmark{1},
    Max~Taylor\IEEEauthorrefmark{2},
    Trevor~Kann\IEEEauthorrefmark{3}, 
    Gary~T.~Leavens\IEEEauthorrefmark{1}, 
    William~L.~Harrison\IEEEauthorrefmark{2},
    and~Lance~Joneckis\IEEEauthorrefmark{2}%
\thanks{\IEEEauthorrefmark{1}University of Central Florida}
\thanks{\IEEEauthorrefmark{2}Idaho National Laboratory; Research Acceleration Department}
\thanks{\IEEEauthorrefmark{3}Carnegie Mellon University}
}

\maketitle

\begin{abstract}

Unmanned aerial vehicles (UAVs) depend on untrusted software components to automate dangerous or critical missions, making them a desirable target for attacks.
Some work has been done to prevent an attacker who has either compromised a ground control station or parts of a UAV's software from sabotaging the vehicle, but not both.
We present an architecture running a UAV software stack with runtime monitoring and seL4-based software isolation that prevents attackers from both exploiting software bugs and stealthy attacks. 
Our architecture retrofits legacy UAVs and secures the popular MAVLink protocol, making wide adoption possible.


\end{abstract}

\section{Introduction}
Unmanned aerial vehicles (UAVs) constitute a wide range of devices.
UAVs can be used to perform a number of tasks, such as those that are too dangerous or costly for a manned aerial vehicle~\cite{greyb2024precision}, or can automate large-scale tasks~\cite{amazonprimeair}. 
To execute these tasks, UAVs communicate with a ground control station (GCS) that sends the UAV commands and receives data from it.
Processing these commands requires sophisticated flight control software (FCS), which needs additional support software, such as the Linux operating system.
This provides a large surface area for attackers; these attacks could crash the UAV or execute an alternative mission~\cite{moder_drone_sok, drone_sensor_sok}.
Furthermore, current communication protocols between the GCS and UAV do not consider the effects of protocol misuse (by an adversary or unskilled pilot) and often have implementation bugs, giving more avenues for an adversary to compromise the UAV~\cite{nasafm25, MissionPlannerIssue1248, RVFuzzer}. 
Therefore, it is important that UAVs are secured against a number of attack vectors.

DARPA's High-Assurance Cyber Military Systems program (HACMS) showed how verified software successfully prevented attackers from exploiting software bugs to take over a UAV~\cite{hacms}.
By isolating software within the UAV, researchers prevented bugs in unverified software from affecting critical software components, such as the FCS~\cite{Collins2017}.

Although these results are promising, HACMS did not consider how sending seemingly benign commands to the UAV could result in failures; i.e., stealthy attacks.
\textit{Stealthy attacks} exploit commands allowed by the communication protocol that maliciously induce undesired UAV behavior~\cite{stealthyCPS,stealthyICS}.
An example of such an attack is an adversary commanding that a UAV's parachute be opened while it is ascending, since current FCSs do not first check to see if parachute-opening preconditions are met. 

Research on runtime monitoring, such as DATUM~\cite{nasafm25}, addresses these protocol-level issues.
DATUM uses a runtime monitor to verify that messages going from the GCS to the UAV follow their intended use and obey rules about message order and parameter values.  
These runtime checks successfully prevent attackers from exploiting stealthy attacks when DATUM is verifying incoming messages. 
However, UAVs often employ less-tested, unverified code such as network drivers~\cite{networkStackVulnerabilities, CVE-2022-41674, CVE-2022-42719, CVE-2022-42720} or computer vision software~\cite{clement_uav_vision}. 
As HACMS showed, an adversary capable of exploiting these vulnerabilities is able to redirect information flow within the drone and bypass DATUM's runtime checks. 

In this paper, we show how to compose contributions from HACMS and DATUM together in a single architecture to defend against a more powerful adversary than any of these prior works alone. 
Like HACMS, our approach isolates software partitions in the UAV, preventing potential vulnerabilities in untrusted software from affecting the FCS while still allowing pre-defined, monitored
communication between partitions. 
Like DATUM, our approach monitors commands sent to the FCS to prevent stealthy attacks.
Our approach is applicable to a variety of available UAVs and compatible with the MAVLink protocol, a popular FCS protocol.

We will first discuss background on UAV security and further explain the threat models of previous works and why they are incompatible (Section~\ref{sec: background}).
Then, we discuss some of the attacks that are not prevented by current UAV security measures (Section~\ref{sec: case studies}).
We show how to defend against these attacks with our system architecture, including nontrivial engineering contributions such as how to preserve communication channels in isolated system partitions (Section~\ref{sec: methodology}).
Finally, we demonstrate that our architecture is practical and effective (Section~\ref{sec: eval}) before discussing potential limitations and concluding (Sections~\ref{sec: discussion} \&~\ref{sec: conclusion}).

Ultimately, our contributions enhance the security and resiliency of UAVs by:
\begin{itemize}
    \item Retrofitting a UAV running widely used FCS to use an architecture that performs runtime monitoring and effectively isolates unverified components, such as the network stack, to ensure FCS integrity and prevent stealthy attacks.
    \item Presenting three case studies that show how this architecture mitigates stealthy attacks, and
    \item  Experimentally demonstrating that the overhead of this approach is acceptable.  
\end{itemize}

\restoregeometry
\section{Background}
\label{sec: background}

In this section, we first elaborate on the HACMS UAV security project (Section
\ref{hacms}) followed by an introduction to the MAVLink protocol and its security flaws (Section \ref{MAVLink}). 
Then we discuss how runtime verification mitigates some security shortcomings of MAVLink while being more accessible than statically verified software (Section \ref{Runtime}).
Finally, we introduce an attack model that poses a threat to both the HACMS approach and runtime monitoring alone (Section \ref{Threat}).

\subsection{The HACMS project}
\label{hacms}

DARPA began the HACMS project to use high-assurance software to better secure cyber-physical systems, including UAVs~\cite{hacms}. 
Undefended UAVs were shown to be hijackable by an adversary exploiting network drivers~\cite{Collins2017}.
Researchers showed how seL4~\cite{sel4}, a formally verified microkernel, prevented exploits in one piece of software from affecting another. 

To further increase UAV resiliency, researchers developed SMACCMPilot, a high-assurance autopilot software that ensured incoming commands and payloads were well-formed~\cite{smaccmpilot},
i.e., all commands were authenticated and conformed to the communication protocol, meaning malformed messages could not be leveraged to compromise the FCS.
However, this guarantee only mitigated the effects of malformed messages: if a protocol allows a well-formed message to maliciously affect the UAV, HACMS and SMACCMPilot would be vulnerable~\cite{nasafm25}.
Despite its formal guarantees, SMACCMPilot has not seen widespread adoption.

\subsection{MAVLink Challenges}
\label{MAVLink}
MAVLink is a widely supported protocol used in GCS-UAV communications~\cite{mavlink}.
MAVLink defines over 300 commands with a wide variety of effects, including commands that change the flight path, dynamically alter flight-control parameters, or trigger actions like deploying a parachute.

Despite its wide adoption, MAVLink has many known safety and security shortcomings~\cite{allouch2019mavsec,hamza2024mavlink,kwon2018empirical,taylor2021study}.
Notably, prior research showed how seemingly benign MAVLink commands caused the UAV to crash or exhibit undesired behavior~\cite{RVFuzzer, kim2021pgfuzz}.
These attacks are known as stealthy attacks~\cite{stealthyICS} and are extremely difficult to detect, even for modern runtime monitoring techniques~\cite{stealthyCPS}, provenance analysis~\cite{provenance}, and physics-based anomaly detection~\cite{pbad1,pbad2,pbad3}.
Widely used UAV FCS, like ArduPilot~\cite{ardupilot} and PX4~\cite{px4}, implement MAVLink and inherit its vulnerabilities.

\subsection{Runtime Monitoring}
\label{Runtime}

Successful runtime monitoring prevents FCSs from accepting commands that violate safety properties while allowing safe commands~\cite{protoARMA, uasVerification, runtimeMon1 ,RuntimeMon2,icss24}.
Notably, DATUM~\cite{nasafm25} specifically targets stealthy attacks.
DATUM augments existing protocols with additional constraints and checks that commands conform to those constraints as they are received. 

Before DATUM, the only way to fix stealthy attack exploits was to report them and wait for control software engineers to patch them.
Worse yet, although there are tools capable of finding such vulnerabilities, researchers report that few of these vulnerabilities are even considered problematic by software maintainers, and thus 
few are ever patched~\cite{RVFuzzer, kim2021pgfuzz}.
DATUM allows operators to immediately retrofit their UAVs, writing constraints on commands and payloads to mitigate unpatched vulnerabilities.

DATUM showed early results in preventing functional manipulation of the MAVLink protocol \cite{nasafm25}. 
However, the use of DATUM relies on its ability to check commands before they are given to the FCS, so the use of DATUM assumes that an attacker cannot bypass these checks, an assumption at odds with the established HACMS threat model~\cite{hacms}.
Such changes would still allow stealthy attacks.

\begin{figure*}
    \centering
    \includegraphics[width=1\linewidth]{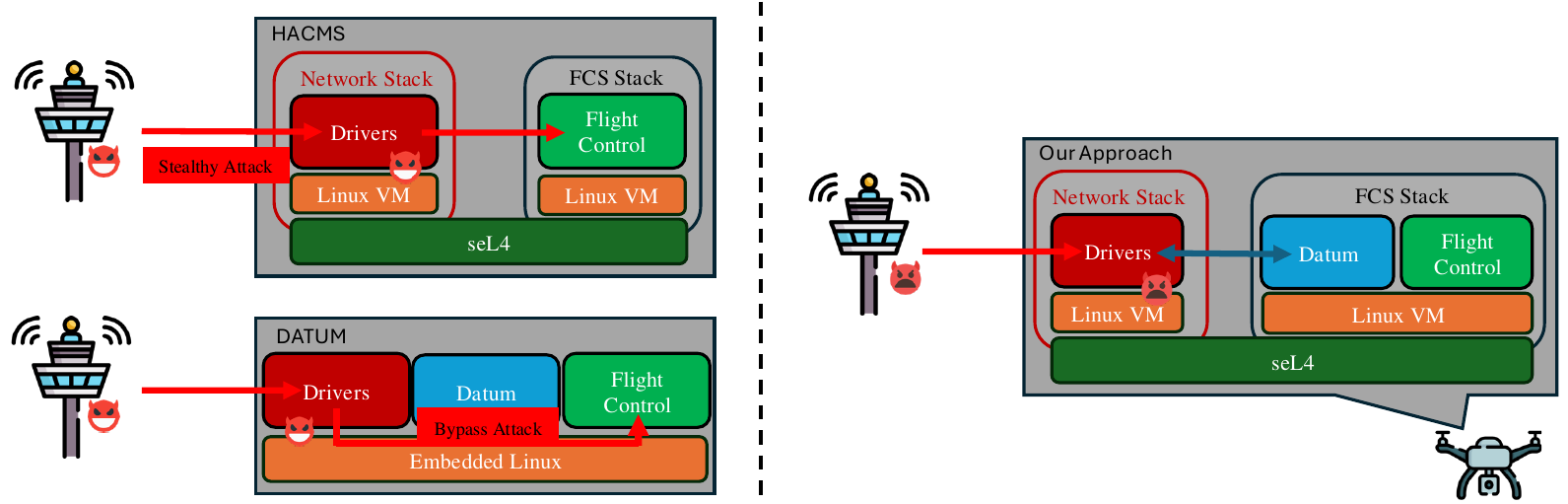}
    \caption{The goal of our approach is to retrofit legacy flight control software to use DATUM in a HACMS-style architecture.
    HACMS successfully prevents an adversary from gaining control of a UAV by compromising unverified software but fails to address stealthy attacks. 
    Dynamic verification approaches, like DATUM, prevent stealthy attacks from compromising a UAV but do not address software vulnerabilities.
    Our approach uses the strength of both of these architectures to prevent a variety of attacks against compromised software and faulty protocols.}
    \label{fig:Goalpic}
\end{figure*}

\subsection{Threat Model}
\label{Threat}
The HACMS architecture defends against compromised drivers but fails to prevent an attacker (e.g., a disgruntled employee) from mounting a stealthy attack on a UAV (Figure~\ref{fig:Goalpic}'s HACMS box).
Conversely, DATUM defends against stealthy attacks but remains vulnerable to an attacker who compromises a UAV's drivers (Figure~\ref{fig:Goalpic}'s DATUM box).
Hence, a resourceful attacker that has compromised both the GCS \emph{and} a UAV driver is able to sabotage a UAV against either standalone approach.

Our threat model assumes that an attacker can compromise the GCS and can send malicious, well-formed, and authenticated commands to the UAV.
The attacker can also compromise some software on the UAV other than the FCS. 
Linux network stack implementations have been shown to introduce vulnerabilities in the past, such as the \emph{challenge ACK} Linux TCP stack vulnerability \cite{2016LinuxStack}, which enabled off-path attacks.
For this paper, we assume an attacker can exploit some vulnerability in the Linux wireless stack (e.g., ~\cite{CVE-2022-41674,CVE-2022-42719,CVE-2022-42720}) to compromise the Linux network drivers. 
In short, a UAV must address an adversary that can both compromise untrusted software and conduct stealthy attacks .

\section{Case Studies}
\label{sec: case studies}

In this paper, we focus on three subtypes of stealthy attacks that threaten UAVs: inaccurate bounds, policy violations, and resource misusage attacks.
In this section, we describe each exploit and its impact.
In Section \ref{sec: methodology}, we show how our approach mitigates their effects.

\subsection{Case Study I: Inaccurate Bounds}
\subsubsection{Background}
Inaccurate bounds attacks exploit wrongly specified or misimplemented command-payload-bounds to put a UAV at risk. 
After determining that a command is well-formed, the FCS checks if the command's payload is within the specified bounds.
If the bounds are not specified correctly, then certain allowed bounds will put the UAV in an at-risk state, where minor changes in external factors, such as a gust of wind, cause the UAV to lose control.
Note that this type of exploit does not immediately cause bad behavior.

The PX4~\cite{px4} autopilot software has 200 UAV parameters that the GCS can dynamically adjust with MAVLink commands. 
The range of safe parameter values sometimes depends on other parameter values.
PX4's pitch parameters are one example. 
\mbox{\lstinline{MC_PITCH_P},} \lstinline{MC_PITCHRATE_FF}, and \lstinline{MC_PITCHRATE_P} are used to adjust the UAV's pitch controls in response to external factors like wind.
Table~\ref{tab:px4_parameters} shows their usage and specified bounds.

\begin{table}[h]
\caption{\rm Select PX4 Controller Parameters and Descriptions}
\centering
\scriptsize
\begin{tabular}{|c|c|c|c|}
\hline
\textbf{Parameter} & \textbf{Description} & \textbf{Max} & \textbf{Default} \\
\hline
\textbf{MC\_PITCH\_P} & Pitch proportional gain & 12 & 6.5 \\
\hline
\textbf{MC\_PITCHRATE\_P} & Pitch rate proportional gain & 0.6 & 0.15 \\
\hline
\textbf{MC\_PITCHRATE\_FF} & Pitch rate feedforward & n/a & 0.0 \\
\hline
\textbf{MC\_PITCHRATE\_MAX} & Max pitch rate limit (deg/s) & 1800.0 & 220.0 \\
\hline 
\end{tabular} 
\label{tab:px4_parameters}
\end{table}

\subsubsection{Impact}
The pitch parameters' specified bounds do not consider interdependencies between the parameters.
More precisely, Kim et al. showed that, when the values of \lstinline{MC_PITCH_P} and \lstinline{MC_PITCHRATE_FF} are high, a pitch rate value of 221 deg/s caused the UAV to crash~\cite{RVFuzzer}.
The only way to constrain the pitch rate in PX4 is with the \mbox{\lstinline{MC_PITCHRATE_MAX}} $n$ payload . 
However, that payload itself is dynamically adjustable, with an allowed upper bound of 1800 deg/s.
Therefore, an attacker can leverage a sequence of valid \lstinline{PARAM_SET} commands to sabotage the UAV.

\subsection{Case Study II: Precondition Violation}
\subsubsection{Background}
The MAVLink protocol also allows commands to be executed that immediately cause harm to the UAV. 
ArduPilot's \lstinline{MAV_CMD_DO_PARACHUTE(2)} command deploys the UAV's parachute.
However, releasing a parachute at the wrong time might jeopardize the UAV's integrity.

ArduPilot's documentation describes the intended preconditions for the \mbox{\lstinline{MAV_CMD_DO_PARACHUTE(2)}} command:
\begin{itemize}
\label{sec: pre conditions}
    \item The motors are armed.
    \item The UAV is in any flight mode but FLIP or ACRO.
    \item The barometer shows the UAV is not gaining altitude.
    \item The UAV's altitude is above the \lstinline{CHUTE_ALT_MIN} parameter's defined altitude.
\end{itemize}

\subsubsection{Impact}
Kim et al. demonstrated that such preconditions are not enforced at runtime~\cite{kim2021pgfuzz}. 
An attacker was able to deploy the parachute while ascending, violating these preconditions. 
Thus, an attacker who is capable of issuing this command puts the UAV at risk.
Currently, there is no mechanism in MAVLink to enforce preconditions.

\subsection{Case Study III: Resource Misusage}
\subsubsection{Background}
Resource misusage 
alters a UAV's mission directives.
Mission directives are important to ensure UAV safety while performing a mission, as they include geofences and safepoints.
The GCS sends these waypoints to the UAV using a \mbox{\lstinline{MISSION_COUNT(N)}} command. 
The initial payload $N$ represents how many waypoints the UAV should expect from the GCS.
Some implementations of the MAVLink protocol expect that the GCS will send exactly $N$ waypoints, leading to undefined behavior when the UAV receives a number of waypoints different than $N$.

A previously reported incident~\cite{MissionPlannerIssue1248}
highlights this vulnerability in a MAVLink mission management implementation.
In this incident, an unsuspecting UAV pilot resent waypoints that triggered errors.
These errors caused the pilot to send a second mission to the UAV.
This second mission exhibited undesired behavior: the drone suddenly stopped behaving as instructed by the second mission and instead started behaving similarly to the first mission.
This was caused by failed mission directive uploads, which were stored in a buffer and only processed once the initial mission ended, causing the UAV to reuse waypoints that were assumed (by the pilot) to have been discarded.

\subsubsection{Impact}
Such unexpected behavior can be catastrophic, as a UAV following a previous mission's waypoint may result in that UAV entering unintended airspace or colliding due to altered geofences.
This behavior was only possible because the user was allowed to input more than the specified $N$ waypoints for the first mission; 
additionally, the UAV incorrectly accepted more than $N$ waypoints for the second mission.
An attacker might leverage misimplemented  resource usage by storing up old mission directives to later trigger them to cause a critical mission disruption.
\section{Methodology}
\label{sec: methodology}
HACMS and DATUM are complimentary approaches to UAV resiliency: they both solve different problems but fail to consider some of what the other does. 
The HACMS model assumes only malformed messages trigger undesired UAV behavior.
However, stealthy attacks violate this assumption by exploiting well-formed commands of overly permissive protocols~\cite{RVFuzzer, kim2021pgfuzz,stealthyCPS,stealthyICS}.
Likewise, DATUM assumes all incoming commands will be vetted by DATUM before the FCS processes them.
However, an adversary with root access to the UAV's OS could redirect incoming packets directly to the FCS, bypassing DATUM.
Thus, we find it effective to combine both resiliency approaches in a single architecture to defend against a stronger adversary.

Our approach retrofits legacy UAVs in a HACMS-style architecture while DATUM checks commands at runtime to prevent stealthy attacks.
This mitigates the risks posed by sophisticated adversaries who have compromised the GCS and then mount a stealthy attack on untrusted UAV software.
In our system, seL4 isolates the UAV's FCS stack from the networking stack. 
Even though the attacker can still issue malicious messages to the FCS, DATUM's runtime monitoring will prevent such messages from reaching the FCS.
Software isolation ensures DATUM will always check messages the GCS sends. 
This architecture is non-trivial: UAV software components must be isolated in seL4 but still be able to communicate in a safe manner, we discuss how to overcome this challenge in section \ref{sec: iso comms}.

\subsection{Architecture}
\begin{figure}
    \includegraphics[width=.9\linewidth]{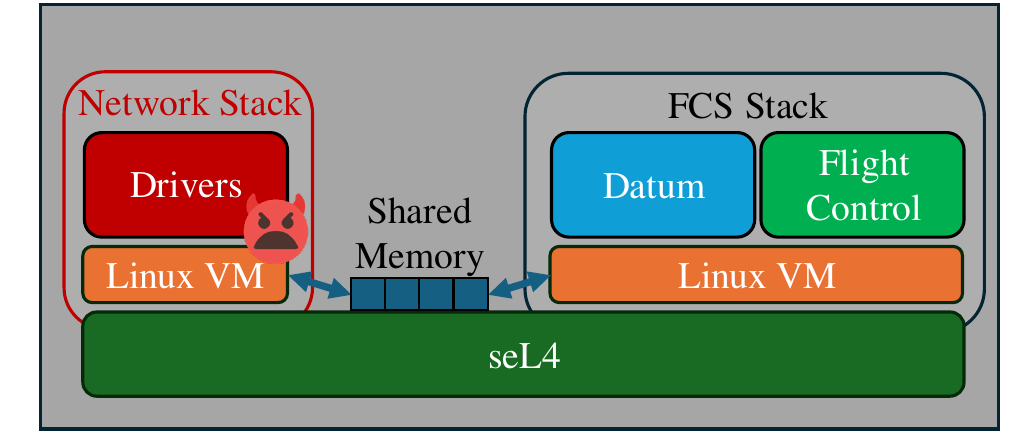}
    \caption{Our approach isolates the network stack from DATUM and the FCS.
This ensures DATUM and the FCS receive packets via entirely separate mechanisms in the Linux kernel.
A bug in one kernel mechanism will affect other software stacks, protecting DATUM and the FCS from compromise.}
    \label{fig:arch}
\end{figure}

Our goal is to augment the HACMS approach by preventing an adversary, who has compromised the GCS and the UAV's networking stack, from conducting a stealthy attack.
In this section, we detail how we accomplish this goal by combining HACMS and DATUM.  
First, we prevent potential vulnerabilities in the networking stack (or other non-verified software) from affecting the FCS by isolating it from the rest of the system (Section~\ref{sec: iso components}).
However, the networking stack cannot be completely separated since we must allow appropriate data from the network (i.e., GCS commands) to contact the FCS; so specific, safe data must still be able to be sent (Section~\ref{sec: iso comms}).
Finally, we detail some additional tweaks our architecture requires to ensure all components integrate together correctly (Section~\ref{sec: integrating}). 

\subsubsection{Isolating System Components}
\label{sec: iso components}
\paragraph{Challenge}
Vulnerabilities reside in the software systems of our dependencies, such as Linux and its drivers. 
This software is difficult to implement and often has errors, but must be relied on. 
To continue with our running story, we will use network drivers as representative of any untrusted software.
To utilize untrusted network drivers, the entire networking stack must be isolated from the rest of the system.

% ...

\paragraph{Approach \& Implementation}
We partition the system into a FCS stack and a networking stack using seL4~\cite{sel4}, as shown in \autoref{fig:arch}.
seL4 has been formally proven to prevent components from interfering with each other.
However, seL4 is difficult to use, as it lacks out-of-the-box support for even basic features like file systems.
To ease this difficulty, we use Microkit~\cite{microkit}. 
Microkit provides a general purpose framework for configuring \textit{protection domains}, which are Microkit's isolated, process-like objects.

Systems built with Microkit have configurations contained in a system description file.
The description file provisions the system's memory region with names, sizes, and physical addresses.
Notably, we map a memory region named \lstinline{ethernet} directly onto QEMU's virtio-net physical memory region.
Similarly, we map a memory region named \lstinline{uart} directly onto a PL011 device.

The system description file maps memory regions into protection domains. 
Memory regions not explicitly mapped into a protection domain are inaccessible to it.
This property is formally proven~\cite{sel4}, providing a high degree of assurance that it is strictly enforced.
Our system description file's protection domains  do not permit the networking stack to access the FCS stack's memory or thread control mechanisms.
Therefore, the FCS stack's data and control flow integrity are not impacted by a possibly exploited networking stack. 

\subsubsection{Enabling Safe Communications Between Components}
\label{sec: iso comms}
\paragraph{Challenge} With the networking stack isolated from the FCS stack, the FCS requires a new mechanism to receive network data.
This mechanism should bypass vulnerabilities that could lead to the FCS stack becoming compromised.
Simultaneously, it should avoid demanding modifications to the FCS stack to maintain compatibility.
Thus, networking must appear unmodified to the FCS in our system.   

\paragraph{Approach \& Implementation}
Our key insight is that vulnerabilities may arise from three sources in the TCP/IP model: the data link layer, the network layer, or the transport layer.
The data link layer may contain vulnerabilities due to bugs in drivers.
These bugs are avoided by using virtual network devices.
In the network layer, we avoid vulnerabilities by forbidding the networking stack from producing IP packets.
In the transport layer, we similarly prevent the networking stack from manipulating UDP packets and TCP segments. 
These small design decisions tremendously reduce the attack surface.

In our approach, network data reaches the FCS stack via memory shared with the networking stack. 
This is shown in \autoref{fig:arch}.
The networking stack decodes MAVLink messages and writes them to the shared memory buffer.
The message is read from shared memory by a process in the FCS stack and forwarded to the FCS. 
%
Overall, an attacker who has compromised software in the unverified stack is unable to influence routing decisions and session management.  

However, application-level vulnerabilities remain exploitable via stealthy attacks. 
Messages exchanged between the GCS and UAV are checked by DATUM.
We use DATUM to refine protocols to defend against stealthy attacks.
DATUM is well-suited for this purpose due to its integration with the F* theorem prover~\cite{fstar}. 
This integration provides a pathway to protocol specifications with provable security properties. 

\subsubsection{Integrating the System}
\label{sec: integrating}
\paragraph{Challenge} 
The system we have described so far constitutes a complex whole.
It introduces two virtual machines, two regions of shared memory, application-level network proxies, and online monitoring.
Thus, we need engineering practices that enable rapid testing and validation. 

\begin{figure}
    \centering
    \includegraphics[width=\linewidth]{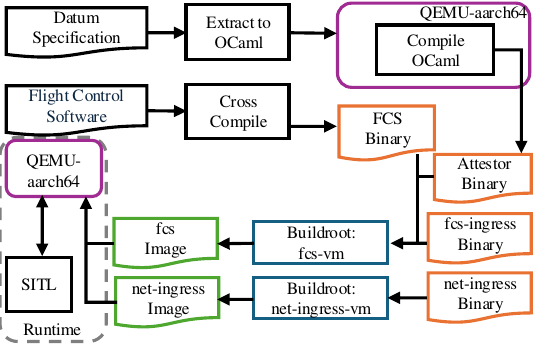}
    \caption{Our system integrates protocol specifications written in DATUM, FCS, and seL4.
The final system is tested and validated using existing SITL infrastructure.}
    \label{fig:integration}
\end{figure}

\paragraph{Approach \& Implementation}
\autoref{fig:integration} shows how the components comprising our system integrate to support rapid testing and validation.
Starting from the top-left corner, the first step is cross-compilation.
DATUM is written in a domain specific language embedded in F*, which supports extraction to executable OCaml.
This extraction must occur on a workstation equipped with an amd64 processor due to a limitation of F*.
The extracted OCaml is compiled in a virtual machine equipped with an aarch64 processor to produce an aarch64 binary.
The FCS, attestor, fcs-ingress (handles communication between shared memory and FCS), and net-ingress (handles communication between shared memory and network interface) binaries are produced by the cross compilation processes.  

The cross-compiled binaries are inputs to virtual machine image builds\footnote{We used Buildroot~\cite{buildroot}, which is a
tool for creating customized embedded Linux distributions, providing out-of-the-box support for dozens of boards, some of which are used in UAVs.
Buildroot produces image files that can execute in a Microkit virtual machine on seL4.}.

We implemented our approach for ArduPilot's and PX4's existing software-in-the-loop (SITL) frameworks.
Importantly, our approach requires no changes to the SITL frameworks.
We provide support in QEMU's ARM ``virt'' virtual machine environment.
This environment resembles compute resources available on UAVs, ensuring our approach is effective across a spectrum of platforms.
Targeting QEMU provides a host of additional benefits, including safe and repeatable testing, easier debugging, and fault injection to test software resilience.

\subsection{Case studies with DATUM}
\label{sec: DATUM case studies}
In Section \ref{sec: case studies}, we defined three types of stealthy attacks and their impacts.
Now, we explain how to mitigate each attack using DATUM in our architecture.
DATUM constrains protocols using the following constructs:
\begin{itemize}
    \item \textbf{Messages:} DATUM describes the commands and payloads that may be sent from each component in the system.
    \item \textbf{Refinements:} Refinements express constraints that are checked at runtime.
    \item \textbf{Constrained Choice:} In situations where a single command can trigger differing command sequences, DATUM uses refinements to express control flow conditions. 
    \item \textbf{Bounded Iteration:} DATUM allows users to write iterative protocols with specified exit conditions and state updates.
\end{itemize}

\subsubsection{Inaccurate bounds}
A dynamic parameter change command should not be accepted just because it is within the range allowed by MAVLink.
Kim et al. emphasize that \lstinline{PARAM_SET(MC_PITCHRATE_MAX,n)} may not be a safe command for a given $n$ because the UAV's safe pitch rate is highly dependent on values of the \lstinline{MC_PITCH_P}, \mbox{\lstinline{MC_PITCHRATE_FF},} and \lstinline{MC_PITCHRATE_P}\cite{RVFuzzer}.
To find out if the value $n$ is safe given the current values of these other parameters, the value of \lstinline{MC_PITCHRATE_MAX} must satisfy an algebraic formula that captures their interdependency.
Such a formula can be expressed as:

\[
\begin{aligned}
n \leq  (m_{1} \times &\texttt{MC\_PITCH\_P}) \times  (m_{2}\times \texttt{MC\_PITCHRATE\_P}) \\
&\times (m_{3} \times \texttt{MC\_PITCHRATE\_FF}) 
\end{aligned}
\]

\hspace*{-0.8ex}where $m_{1}$, $m_{2}$, and $m_{3}$ are adjustable weights associated with each parameter.
Multidimensional fuzzing approaches, such as RVFuzzer\cite{RVFuzzer}, can be used to find suitable values for parameter weights.

Therefore, to dynamically change a parameter, one must first check if dependent parameters have dangerously high or low values, even if they conform to the protocol.
This constraint can be checked by DATUM.
Whenever DATUM receives a parameter-update command, it checks if the command's payload satisfies the constraint and only forwards it to the FCS if it does. 
Inaccurate bounds bugs, such as the pitch parameters exploit, can thus be mitigated using DATUM.

\subsubsection{Precondition Violation}
Precondition violation bugs can be mitigated in a similar way, by checking that the UAV's state satisfies the precondition. For example, 
a \lstinline{MAV_CMD_DO_PARACHUTE(2)} command must be accompanied by a verifiable boolean condition that states such a precondition.
An adversary then cannot incorrectly trigger a parachute release.

\subsubsection{Resource Misusage}
An iterative protocol, such as MAVLink's missions subprotocol, can be defined using DATUM. 
However, it important to first specify the initiating command, which would be  \lstinline{MISSION_COUNT$(N)$}.
The theory underlying DATUM protocols is beyond the scope of this paper (but see the DATUM paper~\cite{nasafm25} for details).

DATUM enforces the constraint that the GCS must send exactly $N$ waypoints without repetition.
DATUM is used to prevent an adversary from exploiting a resource misusage bug to cause undesired UAV behavior in a future flight as in the incident discussed above~\cite{MissionPlannerIssue1248}.

\section{Evaluation \& Results}
\label{sec: eval}

We evaluated our approach using the latest versions of ArduPilot and PX4 (4.5.7 and 1.15.3 respectively) in their existing SITL infrastructure. 
For ArduPilot's evaluation, we used its built-in simulator.
For PX4's evaluation, we used jMAVSim\footnote{We could, in principle, use the more robust Gazebo simulation framework~\cite{gazebo}. 
Gazebo is the preferred simulation environment of PX4. However, for ease of implementation, we elected to use a MAVLink-based simulator.}.  
Both projects were evaluated in a QEMU virtual machine that was hosted on a machine running Ubuntu 24.04 and equipped with 64 GB of memory and a 16-core Intel processor. 

We considered three configurations in our evaluation: QEMU-Linux, QEMU-HACMS, and QEMU-HACMS-DATUM. 
The QEMU-Linux configuration consists of the flight control software executing in a Linux environment on QEMU. QEMU-HACMS uses seL4 to isolate the networking stack from the flight control software. 
QEMU-HACMS-DATUM uses DATUM to detect stealthy attacks.

In each configuration, the flight control software processed between 5,000 and 10,000 MAVLink messages. These messages include sensor readings generated by ArduPilot's and PX4's MAVLink simulation modules. They also include ordinary commands used to guide the UAV along its mission. The mission in this evaluation is a 25-waypoint circuit taken from ArduPilot's existing \lstinline{CMAC-copter-navtest}.

\begin{table}[htbp]
    \caption{\rm Performance results, including latency's $95\%$ confidence intervals. The QEMU-Linux platform features the FCS running on Linux. The QEMU-seL4 platform runs the FCS on a Linux VM in seL4, alongside the isolated network stack. The QEMU-seL4-DATUM platform includes DATUM.}
    \centering
    \scriptsize
    \begin{tabular}{|c|c|c|c|}
    \hline
    \textbf{Platform} & \textbf{FCS} & \textbf{Memory}(KB) & \textbf{Latency}(ms) \\
    \hline
    \multirow{2}{*}{QEMU-Linux} & ArduPilot & 38,520 & $3.420 \pm 0.340$ \\
    & PX4 & 89,676 & $1.011 \pm 0.062$ \\  
    \hline

    \multirow{2}{*}{QEMU-HACMS} & ArduPilot & 71,192 & $3.811 \pm 0.750$ \\
    & PX4 & 121,060 & $6.688 \pm 0.193$ \\
    \hline

    \multirow{2}{*}{QEMU-HACMS-DATUM} & ArduPilot & 85,280 & $4.061 \pm 0.619$ \\
    & PX4 & 135,148 & $6.939 \pm 0.191$ \\
    \hline 
    \end{tabular}
    \label{tab:perf-results}
\end{table}

Our evaluation focuses on message latency and memory usage.
Latency is critical for UAVs, as messages must be delivered on time to prevent accidents. Memory usage is important because UAVs are resource-constrained. 

To measure message latency, we added instrumentation to the ground control station, simulation, and flight control software. The instrumentation records the timestamps of each message sent and received. To measure memory overhead, we measured the memory used by each virtual machine on seL4. 
The results of our evaluation are shown in \autoref{tab:perf-results}.

\begin{table}[]
    \caption{\rm We verified the attestor generated by DATUM is able to detect the stealthy attacks described in our case studies.}
    \centering
    \begin{tabular}{|c|c|}
        \hline
         \textbf{Attack} & \textbf{Detected by Attestor?}  \\
         \hline
         Inaccurate Bounds (Case Study I) & Yes \\
         Precondition Violation (Case Study II) & Yes \\
         Resource Misusage (Case Study III) & Yes \\
         \hline
    \end{tabular}
    \label{tab:DATUM-attacks-detected}
\end{table}

To further evaluate our work, we verified that the attestor generated by DATUM is able to detect the stealthy attacks described in our case studies. This is shown in \autoref{tab:DATUM-attacks-detected}. To perform this experiment, we sent MAVLink messages comprising each attack to the attestor generated by DATUM. We checked that the attestor correctly identified each attack. 

We note the increase in latency  while moving from QEMU-Linux to QEMU-HACMS. Despite this overhead, the UAV is still reliably controllable from the GCS. Moreover, the latency is reducible by a more careful implementation of our data exchange scaffolding. 
\section{Discussion}
\label{sec: discussion}

In this section, we  briefly discuss some of the known limitations of our work and their possible fixes as well as potential extensions of this work. 
 
\subsection{The Gap Between Simulation \& Implementation}
Our results are obtained using a simulator and SITL.
Therefore, the latency, memory, and behavioral results may differ in a physical UAV.
However, numerous similar studies have relied on simulations in the past~\cite{taylor2021study,kim2021pgfuzz,RVFuzzer,avis,sa4u}.
We expect our results to be similar on a physical UAV because simulated and physical UAVs share a significant amount of source code.


\subsection{Runtime Performance}
Our architectural changes increase the overhead of processing network requests and executing actions from the autopilot software. 
However, this overhead does not interfere with typical UAV operations. 
Nonetheless, we still briefly discuss overhead here.

The biggest impact on performance is attributed to our current networking approach using a ring buffer to enable communication between Linux virtual machines in order to mitigate data link layer attack vectors (see Section \ref{sec: iso comms}).
However, our approach relies on busy-waiting, which can have direct effects on communication overhead.
It is possible that a more carefully tuned version of our implementation may achieve better results; 
But, to achieve significant performance improvement while maintaining the same level of security, one would have to invoke seL4 system calls from within the Linux VMs.
Such calls are safe, as part of the seL4 proof, and do not require busy-waiting.
This approach requires modifications to the Linux kernel. We leave this for future work.

Additionally, DATUM's type checking at runtime incurs additional overhead~\cite{nasafm25}; however, it is relatively small. 
seL4 does not have a significant overhead~\cite{sel4}. 

\subsection{Human Performance Improvement Approach}
Our approach adds constraints to protocols that prevents commands and payloads that may be used maliciously in stealthy attacks.
However, these commands could also be used by a skilled operator to perform advanced maneuvers while retaining control of the drone.
For example, a skilled operator may trigger a parachute while violating a precondition to reduce the forward speed of the drone and avoid a head-on crash.

Although there may be operators that can perform such feats, there are also unskilled operators that may sabotage a UAVs unintentionally. 
Our work takes a Human Performance Improvement (HPI) approach~\cite{HPI1,HPI2}; we recognize that human error cannot be fully mitigated. Hence, we place controls, in the form of DATUM runtime verification checks, to mitigate the effects of errors.
We also note that DATUM is flexible and allows for engineers to adjust protocol constraints to fit their respective experience and security needs.
\section{Conclusion}
\label{sec: conclusion}

In this work, we have proposed a state-of-the-art threat model and shown how to defend against it using an ensemble of prior work that creates a flexible system for secure UAVs.
We used the formal guarantees of HACMS' architecture to isolate untrusted parts of the UAV's software stack from trusted, critical components. 
We also utilized DATUM's flexible runtime monitoring guarantees to ensure stealthy attacks cannot be executed and to enable engineers to quickly patch protocols. 
Combining these two architectures required non-trivial reengineering of the UAV's communication infrastructure and repackaging commercial UAV software to fit into the HACMS architecture. 
The overhead of our design is nontrivial but small enough to maintain nominal UAV piloting; further optimizations would bring this number down considerably if latency is critical for an application. 
We demonstrate that our system is robust against our new threat model through three case studies.

\bibliography{bib}

\begin{thebibliography}{10}

\bibitem{greyb2024precision}
GreyB, ``Understanding the scope of uavs \& drones for precision agriculture in 2024,'' 2024.
\newblock Accessed: 2025-02-05.

\bibitem{amazonprimeair}
Amazon, ``Prime air: E-commerce foundation.''
\newblock Accessed: 2025-02-05.

\bibitem{moder_drone_sok}
B.~Nassi, R.~Bitton, R.~Masuoka, A.~Shabtai, and Y.~Elovici, ``Sok: Security and privacy in the age of commercial drones,'' in {\em 2021 IEEE symposium on security and privacy (SP)}, pp.~1434--1451, IEEE, 2021.

\bibitem{drone_sensor_sok}
Y.~Xu, X.~Han, G.~Deng, J.~Li, Y.~Liu, and T.~Zhang, ``Sok: Rethinking sensor spoofing attacks against robotic vehicles from a systematic view,'' in {\em 2023 IEEE 8th European Symposium on Security and Privacy (EuroS\&P)}, pp.~1082--1100, 2023.

\bibitem{nasafm25}
A.~Amorim, M.~Taylor, G.~T. Leavens, B.~Harrison, L.~Joneckis, and T.~Kann, ``Enforcing mavlink safety \& security properties via refined multiparty session types,'' 2025.

\bibitem{MissionPlannerIssue1248}
ArduPilot, ``Missionplanner issue \#1248: Add support for new flight modes.'' \url{https://github.com/ArduPilot/MissionPlanner/issues/1248}, 2024.
\newblock Accessed: 2024-11-22.

\bibitem{RVFuzzer}
T.~Kim, C.~H. Kim, J.~Rhee, F.~Fei, Z.~Tu, G.~Walkup, X.~Zhang, X.~Deng, and D.~Xu, ``{RVFuzzer}: Finding input validation bugs in robotic vehicles through {Control-Guided} testing,'' in {\em 28th USENIX Security Symposium (USENIX Security 19)}, (Santa Clara, CA), pp.~425--442, USENIX Association, Aug. 2019.

\bibitem{hacms}
K.~Fisher, J.~Launchbury, and R.~Richards, ``The hacms program: using formal methods to eliminate exploitable bugs,'' {\em Philosophical Transactions of the Royal Society A: Mathematical, Physical and Engineering Sciences}, vol.~375, no.~2104, p.~20150401, 2017.

\bibitem{Collins2017}
R.~Collins and S.~Copy, ``Secure mathematically-assured composition of control models approved for public release; distribution unlimited air force materiel command,'' 2017.

\bibitem{stealthyCPS}
T.~Sui, Y.~Mo, D.~Marelli, X.~Sun, and M.~Fu, ``The vulnerability of cyber-physical system under stealthy attacks,'' {\em IEEE Transactions on Automatic Control}, vol.~66, no.~2, pp.~637--650, 2021.

\bibitem{stealthyICS}
D.~I. Urbina, J.~A. Giraldo, A.~A. Cardenas, N.~O. Tippenhauer, J.~Valente, M.~Faisal, J.~Ruths, R.~Candell, and H.~Sandberg, ``Limiting the impact of stealthy attacks on industrial control systems,'' in {\em Proceedings of the 2016 ACM SIGSAC Conference on Computer and Communications Security}, CCS '16, (New York, NY, USA), p.~1092–1105, Association for Computing Machinery, 2016.

\bibitem{networkStackVulnerabilities}
M.~Hooper, Y.~Tian, R.~Zhou, B.~Cao, A.~P. Lauf, L.~Watkins, W.~H. Robinson, and W.~Alexis, ``Securing commercial wifi-based uavs from common security attacks,'' in {\em MILCOM 2016 - 2016 IEEE Military Communications Conference}, pp.~1213--1218, 2016.

\bibitem{CVE-2022-41674}
{National Vulnerability Database}, ``{CVE-2022-41674: Vulnerability Summary},'' 2022.
\newblock Accessed: 2025-02-02.

\bibitem{CVE-2022-42719}
{National Vulnerability Database}, ``{CVE-2022-42719: Vulnerability Summary},'' 2022.
\newblock Accessed: 2025-02-02.

\bibitem{CVE-2022-42720}
{National Vulnerability Database}, ``{CVE-2022-42720: Vulnerability Summary},'' 2022.
\newblock Accessed: 2025-02-02.

\bibitem{clement_uav_vision}
R.~Martin, C.~Fung, N.~Keetha, L.~Bauer, and S.~Scherer, ``Targeted image transformation for improving robustness in long range aircraft detection,'' in {\em 2024 IEEE/RSJ International Conference on Intelligent Robots and Systems (IROS)}, pp.~10431--10438, IEEE, 2024.

\bibitem{sel4}
G.~Klein, K.~Elphinstone, G.~Heiser, J.~Andronick, D.~Cock, P.~Derrin, D.~Elkaduwe, K.~Engelhardt, R.~Kolanski, M.~Norrish, {\em et~al.}, ``sel4: Formal verification of an os kernel,'' in {\em Proceedings of the ACM SIGOPS 22nd symposium on Operating systems principles}, pp.~207--220, 2009.

\bibitem{smaccmpilot}
P.~C. Hickey, L.~Pike, T.~Elliott, J.~Bielman, and J.~Launchbury, ``Building embedded systems with embedded dsls,'' in {\em Proceedings of the 19th ACM SIGPLAN international conference on Functional programming}, pp.~3--9, 2014.

\bibitem{mavlink}
{MAVLink Development Team}, ``Mavlink: Micro air vehicle communication protocol,'' 2024.
\newblock Accessed: 2024-11-19.

\bibitem{allouch2019mavsec}
A.~Allouch, O.~Cheikhrouhou, A.~Koub{\^a}a, M.~Khalgui, and T.~Abbes, ``Mavsec: Securing the mavlink protocol for ardupilot/px4 unmanned aerial systems,'' in {\em 2019 15th International Wireless Communications \& Mobile Computing Conference (IWCMC)}, pp.~621--628, IEEE, 2019.

\bibitem{hamza2024mavlink}
M.~A. Hamza, M.~Mohsin, M.~Khalil, and S.~M. K.~A. Kazmi, ``Mavlink protocol: A survey of security threats and countermeasures,'' in {\em 2024 4th International Conference on Digital Futures and Transformative Technologies (ICoDT2)}, pp.~1--8, IEEE, 2024.

\bibitem{kwon2018empirical}
Y.-M. Kwon, J.~Yu, B.-M. Cho, Y.~Eun, and K.-J. Park, ``Empirical analysis of mavlink protocol vulnerability for attacking unmanned aerial vehicles,'' {\em IEEE Access}, vol.~6, pp.~43203--43212, 2018.

\bibitem{taylor2021study}
M.~Taylor, J.~Boubin, H.~Chen, C.~Stewart, and F.~Qin, ``A study on software bugs in unmanned aircraft systems,'' in {\em 2021 International Conference on Unmanned Aircraft Systems (ICUAS)}, pp.~1439--1448, IEEE, 2021.

\bibitem{kim2021pgfuzz}
H.~Kim, M.~O. Ozmen, A.~Bianchi, Z.~B. Celik, and D.~Xu, ``Pgfuzz: Policy-guided fuzzing for robotic vehicles.,'' in {\em NDSS}, 2021.

\bibitem{provenance}
X.~Han, T.~Pasquier, A.~Bates, J.~Mickens, and M.~Seltzer, ``Unicorn: Runtime provenance-based detector for advanced persistent threats,'' in {\em Proceedings 2020 Network and Distributed System Security Symposium}, NDSS 2020, Internet Society, 2020.

\bibitem{pbad1}
H.~Choi, W.-C. Lee, Y.~Aafer, F.~Fei, Z.~Tu, X.~Zhang, D.~Xu, and X.~Deng, ``Detecting attacks against robotic vehicles: A control invariant approach,'' in {\em Proceedings of the 2018 ACM SIGSAC Conference on Computer and Communications Security}, CCS '18, (New York, NY, USA), p.~801–816, Association for Computing Machinery, 2018.

\bibitem{pbad2}
F.~Fei, Z.~Tu, D.~Xu, and X.~Deng, ``Learn-to-recover: Retrofitting uavs with reinforcement learning-assisted flight control under cyber-physical attacks,'' in {\em 2020 IEEE International Conference on Robotics and Automation (ICRA)}, pp.~7358--7364, 2020.

\bibitem{pbad3}
F.~Fei, Z.~Tu, R.~Yu, T.~Kim, X.~Zhang, D.~Xu, and X.~Deng, ``Cross-layer retrofitting of uavs against cyber-physical attacks,'' in {\em 2018 IEEE International Conference on Robotics and Automation (ICRA)}, pp.~550--557, 2018.

\bibitem{ardupilot}
{ArduPilot Development Team}, ``Ardupilot: The open source autopilot,'' 2024.
\newblock Accessed: 2024-11-19.

\bibitem{px4}
{PX4 Development Team}, ``Px4: Open source autopilot for drones,'' 2024.
\newblock Accessed: 2024-11-19.

\bibitem{protoARMA}
J.~T. Slagel, L.~M. White, A.~Dutle, C.~A. Mu{\~n}oz, and N.~Crespo, ``A formal verification framework for runtime assurance,'' in {\em NASA Formal Methods Symposium}, pp.~322--328, Springer, 2024.

\bibitem{uasVerification}
J.~T. Slagel, L.~M. White, A.~Dutle, C.~A. Mu{\~n}oz, and N.~Crespo, ``A verification framework for runtime assurance of autonomous {UAS},'' in {\em 2024 AIAA DATC/IEEE 43rd Digital Avionics Systems Conference (DASC)}, pp.~01--08, IEEE, 2024.

\bibitem{runtimeMon1}
E.~Bartocci, J.~Deshmukh, A.~Donz{\'e}, G.~Fainekos, O.~Maler, D.~Ni{\v{c}}kovi{\'{c}}, and S.~Sankaranarayanan, {\em Specification-Based Monitoring of Cyber-Physical Systems: A Survey on Theory, Tools and Applications}, pp.~135--175.
\newblock Cham: Springer International Publishing, 2018.

\bibitem{RuntimeMon2}
E.~Kang, A.~Ganlath, S.~Mishra, F.~Baiduc, and N.~Ammar, ``Contract-driven runtime adaptation,'' in {\em NASA Formal Methods} (N.~Benz, D.~Gopinath, and N.~Shi, eds.), (Cham), pp.~298--313, Springer Nature Switzerland, 2024.

\bibitem{icss24}
A.~Amorim, T.~Kann, M.~Taylor, and L.~Joneckis, ``{ Towards Provable Security in Industrial Control Systems Via Dynamic Protocol Attestation },'' in {\em 2024 Annual Computer Security Applications Conference Workshops (ACSAC Workshops)}, (Los Alamitos, CA, USA), pp.~120--132, IEEE Computer Society, Dec. 2024.

\bibitem{2016LinuxStack}
A.~Quach, Z.~Wang, and Z.~Qian, ``Investigation of the 2016 linux tcp stack vulnerability at scale,'' {\em Proc. ACM Meas. Anal. Comput. Syst.}, vol.~1, June 2017.

\bibitem{microkit}
Z.~A. Kocsis, M.~Paturel, I.~Subasinghe, T.~Weibel, and G.~Heiser, ``the sel4 microkit.'' draft, 2023.

\bibitem{fstar}
N.~Swamy, C.~Hri{\c{t}}cu, C.~Keller, A.~Rastogi, A.~Delignat-Lavaud, S.~Forest, K.~Bhargavan, C.~Fournet, P.-Y. Strub, M.~Kohlweiss, {\em et~al.}, ``Dependent types and multi-monadic effects in f,'' in {\em Proceedings of the 43rd annual ACM SIGPLAN-SIGACT Symposium on Principles of Programming Languages}, pp.~256--270, 2016.

\bibitem{buildroot}
B.~Developers, ``Buildroot: A simple, efficient and easy-to-use tool to generate embedded linux systems.'' \url{https://buildroot.org/}, 2025.
\newblock Accessed: 2025-02-02.

\bibitem{gazebo}
N.~Koenig and A.~Howard, ``Design and use paradigms for gazebo, an open-source multi-robot simulator,'' in {\em 2004 IEEE/RSJ international conference on intelligent robots and systems (IROS)(IEEE Cat. No. 04CH37566)}, vol.~3, pp.~2149--2154, Ieee, 2004.

\bibitem{avis}
M.~Taylor, H.~Chen, F.~Qin, and C.~Stewart, ``Avis: In-situ model checking for unmanned aerial vehicles,'' in {\em 2021 51st Annual IEEE/IFIP International Conference on Dependable Systems and Networks (DSN)}, pp.~471--483, IEEE, 2021.

\bibitem{sa4u}
M.~Taylor, J.~Aurand, F.~Qin, X.~Wang, B.~Henry, and X.~Zhang, ``Sa4u: practical static analysis for unit type error detection,'' in {\em Proceedings of the 37th IEEE/ACM International Conference on Automated Software Engineering}, pp.~1--11, 2022.

\bibitem{HPI1}
A.~Marker, S.~W. Villachica, D.~Stepich, D.~Allen, and L.~Stanton, ``An updated framework for human performance improvement in the workplace: The spiral hpi framework,'' {\em Performance Improvement}, vol.~53, no.~1, pp.~10--23, 2014.

\bibitem{HPI2}
R.~Humphress and Z.~L. Berge, ``Justifying human performance improvement interventions,'' {\em Performance Improvement}, vol.~45, no.~7, pp.~13--22, 2006.

\end{thebibliography}
\bibliographystyle{ieeetr}
\end{document}